# Molecular dynamic simulation of multicomponent CoCrFeNiMn high-entropy alloy thin film deposition

O.I. Kushnerov[a], S.I. Ryabtsev[a] and V.F. Bashev[b]

[a] Department of Experimental Physics, Oles Honchar Dnipro National University, Dnipro, Ukraine;

[b]Department of Condensed Matter Physics, Dniprovsky State Technical University, Kamianske, Ukraine

**CONTACT** O.I. Kushnerov: kushnrv@gmail.com, Department of Experimental Physics, Oles Honchar Dnipro National University, 72, Nauky av., Dnipro, 49010, Ukraine

The deposition and growth processes of a thin high-entropy CoCrFeMnNi alloy film on an Al (100) substrate were studied using molecular dynamics simulations. The interactions between atoms were described with a set of Morse potentials developed using adopted mixing rules for regular solutions. The total simulation time was 100 ns. A total of 50000 atoms, with a high incident energy of 10 eV, were deposited on the substrate, resulting in a film thickness of approximately 6.1 nm. At the end of the simulation, the simulated film contains face-centred cubic (FCC), body-centred cubic (BCC), hexagonal close-packed (HCP) phases, as well as an amorphous phase. An analysis of the radial distribution function (RDF) of atoms allowed for the determination of the nearest neighbour distances and estimation of the lattice parameters for these phases. The phase composition and crystal structure parameters of the obtained film are in good agreement with experimental data.

## 1. Introduction

Multicomponent high-entropy alloys (HEAs), a novel class of materials comprising multiple principal elements in near-equiatomic or equiatomic ratios, have attracted significant attention due to their exceptional combinations of strength, ductility, thermal stability, and resistance to extreme environments [1-7]. Unlike traditional alloys, which are typically composed of one or two principal elements, HEAs consist of five or more elements. This complex composition leads to a high configurational entropy, which stabilizes the solid solution phase and contributes to the exceptional properties of HEAs. While bulk HEAs have been extensively investigated, their application as thin films introduces additional complexities and opportunities that demand systematic investigation at the atomic scale. This is even more important because HEA thin films have emerged as promising candidates for protective coatings, offering enhanced durability and surface performance for industrial, aerospace, and electronic applications [8-10].

However, the phase formation mechanisms in HEAs fundamentally differ between sputtered thin films and bulk materials due to the distinct kinetic and thermodynamic conditions inherent to the sputtering process. While bulk HEA synthesis typically occurs under near-equilibrium conditions with cooling rates of $10^1$-$10^3$ K/s, the physical vapor deposition (PVD) environment subjects atoms to extreme quenching rates exceeding $10^{12}$ K/s [9]. This ultrafast cooling substantially suppresses atomic diffusion and redistributional processes. This kinetic limitation prevents the system from reaching its thermodynamic equilibrium configuration, leading to the stabilization of metastable phases and trapping of structural defects within the film microstructure. In contrast, bulk HEA synthesis allows greater atomic mobility and the achievement of more stable, equilibrium-like phases.

Among the various methods available for thin-film synthesis, ion-plasma deposition, a PVD method suggests unique advantages for HEA film synthesis [9]. This technique involves the generation of a plasma containing the desired metal atoms, which are deposited onto a substrate. The high-energy atom bombardment inherent to ion-plasma deposition induces unique dynamic effects, including atomic mixing, preferential sputtering, and stress generation, which can significantly impact the structure and properties of the resulting film. Understanding these processes at the atomic scale is critical for optimizing the deposition conditions and achieving films with tailored microstructures and superior performance.

Molecular dynamics (MD) is one of the most effective methods for modelling the structure and properties of materials [11]. MD simulation offers a powerful computational tool for investigating the complex processes involved in ion-plasma deposition of HEA films. By modelling the interactions between atoms at the atomic scale [11], MD simulations can provide insights into the dynamics of film growth, the formation of defects, and the evolution of microstructure.

This study aims to investigate the ion-plasma deposition of CoCrFeNiMn HEA films using molecular dynamics simulations. The choice of interatomic potential used in MD simulations can have a significant impact on the resulting predictions and insights. For the study of the CoCrFeNiMn HEA system, two distinct potentials have been employed in the literature – the 2NN MEAM (modified embedded-atom method) potential developed in [12] and the Lennard-Jones (LJ) type potential [13].

The MEAM potential offers a more sophisticated description of the complex multi-element interactions present in HEAs, incorporating both the embedding energy and pairwise interactions. This approach has been shown to provide a reliable representation of thermodynamic properties, defect energetics, and short-range ordering behaviour in HEA systems. In contrast, the LJ-type potential represents a more simplified, computationally efficient description of the interatomic forces. While the LJ potential may capture certain trends, it lacks the nuanced treatment of the electronic structure effects that are important for more accurate modelling of the properties of multicomponent alloys.

As highlighted in [13], the concept of an "average atom" [14] allows the treatment of an alloy as a homogeneous material, bypassing the need to account for local variations in chemical composition. This approach enables the use of a single simulation to obtain the average properties of a random alloy [14]. Therefore, the random face-centred cubic (FCC) structure of CoCrFeNiMn alloy, where details of the electronic structure controlling the bonding are not crucial, can be described with a certain degree of accuracy by a simple pair potential [13]. However, the LJ potential, with its two adjustable parameters (σ and ε), is insufficient to accurately describe all properties of the alloy and its components. In this study, we employ the Morse potential, which offers greater flexibility through its three tuneable parameters ($D$, α, and $r_0$), allowing for the modelling of more diverse bonding behaviours. We propose a set of Morse potentials to describe all interactions within the CoCrFeNiMn HEA, as well as between the HEA and the aluminium substrate.

## 2. Simulation model and methods

### *2.1. Simulation model*

Classical molecular dynamics simulations were conducted to study the deposition and growth of thin films of CoCrFeNiMn HEA using the Large-Scale Atomic/Molecular Massively Parallel Simulator (LAMMPS, Sandia National Laboratories, USA) [15] with GPU accelerator package [16,17]. The simulation results were visualized and subsequently analysed using the open-source software Open Visualization Tool (OVITO) [18].

The simulation utilized a three-dimensional simulation cell with periodic boundary conditions applied in the horizontal $x$ and $y$ directions, while a free boundary condition was imposed in the vertical $z$-direction to enable surface growth. The substrate was modelled as FCC Al (100) slab with dimensions of 100×100×18 Å, containing 10368 atoms. Aluminium was chosen because, unlike nickel, the substrate of which was considered in [19,20], it is not a component of the CoCrFeNiMn alloy. To prevent displacement of the substrate due to the impact of incoming atoms, the lower atomic layer of the substrate was fixed. An intermediate region of the substrate above this stationary layer was maintained at a temperature of 300 K controlled by a Langevin thermostat, to ensure isothermal growth conditions. The topmost atoms of the substrate formed a Newtonian region with a thickness of 10 Å and were free to move under the influence of the deposited atoms without any thermostatic constraints.

Atoms of Co, Cr, Fe, Ni, and Mn were deposited onto the substrate in an equimolar ratio at intervals of 10 ps, corresponding to an atomic flux density of $5 \cdot 10^{23}$ atoms/cm$^2$·s. In this manner, 10000 atoms of each type were deposited. The deposition rate of 0.5 atoms/ps was chosen to allow sufficient time to dissipate the heat generated during the deposition process, preventing the deposited layer from overheating. Although this deposition rate is significantly higher than experimental fluxes, the resulting properties of the simulated film, including its morphology, structure, and adhesion can closely correspond to experimental trends, as stated in [21,22]. The simulation time step is set to 1 fs ($1 \cdot 10^{-15}$ s). The simulation ran for $1 \cdot 10^8$ steps, corresponding to 100 nanoseconds.

The deposited atoms were given an initial velocity corresponding to a high energy of 10 eV, with an incident angle of 0°. Their initial positions were randomly distributed in the $x$-$y$ plane above the substrate, ensuring they were outside the range of interaction potentials in the $z$-direction. All atoms were deposited from a height of 140–145 Å above the substrate surface. During simulations, the estimated local heating and cooling rates reach up to $10^{15}$ K/s.

### *2.2. Interatomic potential*

The simulation was carried out using the Morse potential, given by the following equation:

$$\phi(r) = D\left[e^{-2\alpha(r-r_0)} - 2e^{-\alpha(r-r_0)}\right], \qquad r < r_c \tag{1}$$

where $r$ is the distance between two atoms, $D$, $\alpha$, and $r_0$ are potential parameters, $r_c$ is the cutoff distance. Here $\alpha$ represents the width of the potential curve, having dimensions of reciprocal distance, while $D$ represents the depth of the potential well at zero temperature, having dimensions of energy. The variable $r_0$ represents the distance at which the potential is minimized.

The parameters of the potentials for pure elements Al, Co, Cr, Fe, and Ni were developed on the base of parameters, obtained in [23-26], with additional fitting to reproduce experimental values of lattice constant $r_{lat}$ and cohesive energy $E_c$ (Tab.1). Regarding manganese, [27] notes that the low-temperature α-Mn phase exhibits a highly complex 58-atom unit cell, complicating the explicit definition of pair interactions. Consequently, the FCC γ-Mn structure was adopted as the ground state, consistent with [27,28]. The values of the lattice constant and cohesive energy for γ-Mn were taken from [27], and the potential was adjusted to reproduce these values. Fig. 1 shows graphs of the obtained potentials.

**Table 1.** The values of the lattice constant and cohesive energy used for fitting the potentials of pure Co, Cr, Fe, Ni, Mn, and Al [23,27,29].

| Element | $r_{lat}$, Å | $E_c$, eV/atom |
|---|---|---|
| Al (FCC) | 4.0495 | 3.39 |
| Co (HCP) | 2.5071 | 4.39 |
| Cr (BCC) | 2.8848 | 4.1 |
| Fe (BCC) | 2.8665 | 4.28 |
| Ni (FCC) | 3.524 | 4.44 |
| γ-Mn (FCC) | 3.6925 | 2.855 |

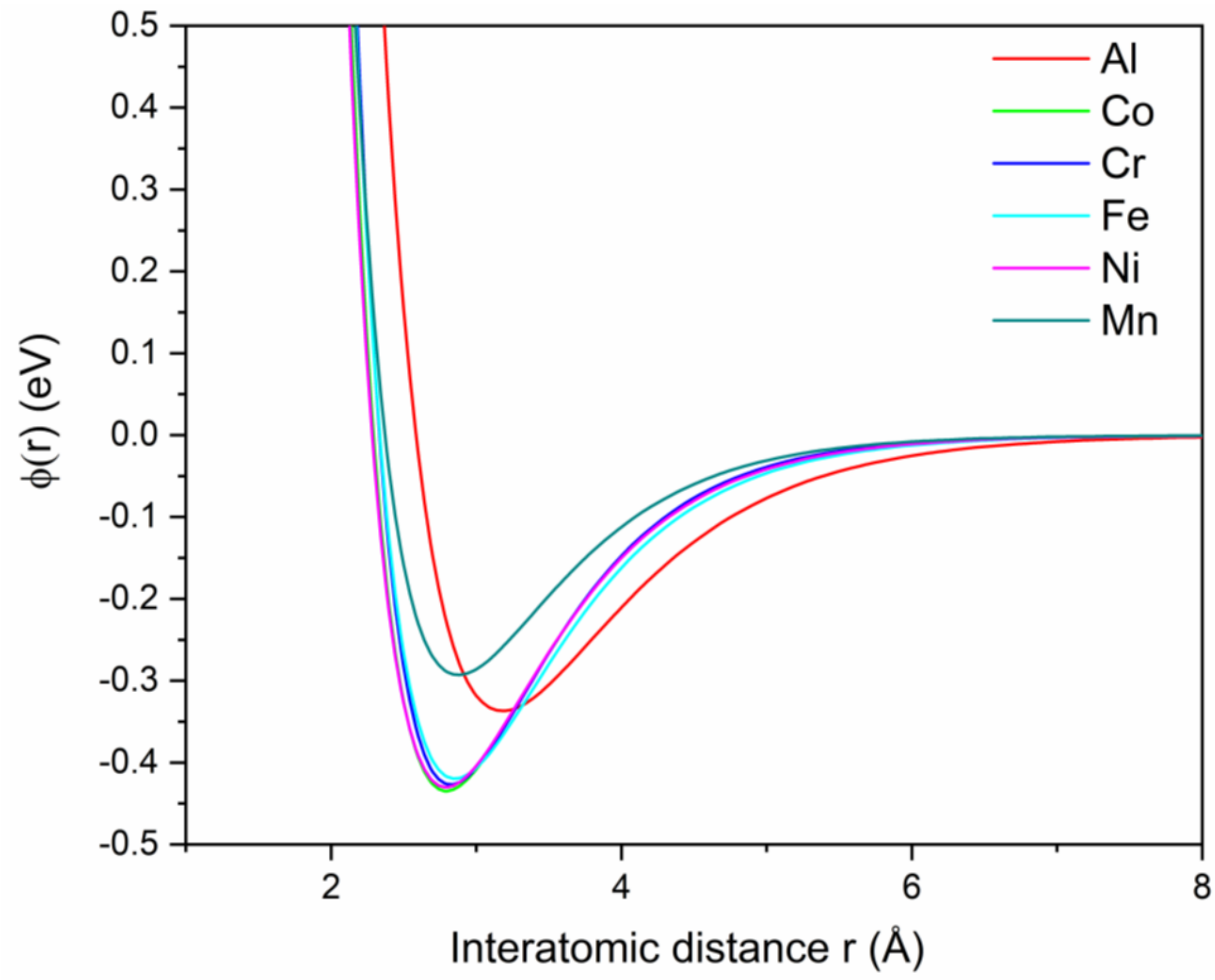

**Figure 1.** Interatomic potential curves for pure Co, Cr, Fe, Ni, Mn, and Al.

The Morse potential parameters $D_{AB}$, $\alpha_{AB}$, and $r_0$ for interatomic interactions between different kinds of atoms A–B were calculated by using empirical combination rules [30]:

$$D_{AB} = (D_A D_B)^{1/2} \quad (2)$$

$$\alpha_{\mathrm{AB}} = \frac{1}{2}(\alpha_{\mathrm{A}} + \alpha_{\mathrm{B}}) \quad (3)$$

$$\sigma_{\mathrm{A}} = r_{0A} - \frac{\ln 2}{\alpha_{\mathrm{A}}} \quad (4)$$

$$\sigma_{\mathrm{B}} = r_{0\mathrm{B}} - \frac{\ln 2}{\alpha_{\mathrm{B}}} \quad (5)$$

$$r_{0AB} = (\sigma_A \sigma_B)^{1/2} + \frac{\ln 2}{\alpha_{AB}} \quad (6)$$

The potential parameters determined in this manner are summarized in Tab. 2.

**Table 2.** Parameters of the Morse potential for Co-Cr-Fe-Ni-Mn-Al system.

| Bond | $D$, eV | α, Å$^{-1}$ | $r_0$, Å |
|---|---|---|---|
| Co-Co | 0.4349 | 1.38 | 2.794 |
| Cr-Cr | 0.4271 | 1.409 | 2.8248 |
| Fe-Fe | 0.4198 | 1.335 | 2.8544 |
| Ni-Ni | 0.4301 | 1.361 | 2.7919 |

| Mn-Mn | 0.293 | 1.3708 | 2.877 |
|---|---|---|---|
| Al-Al | 0.3370 | 1.16 | 3.1823 |
| Co-Cr | 0.431 | 1.3945 | 2.8093 |
| Co-Fe | 0.4273 | 1.3575 | 2.824 |
| Co-Ni | 0.4325 | 1.3705 | 2.7929 |
| Co-Mn | 0.357 | 1.3754 | 2.8352 |
| Co-Al | 0.3828 | 1.27 | 2.9793 |
| Cr-Fe | 0.4234 | 1.372 | 2.8392 |
| Cr-Ni | 0.4286 | 1.385 | 2.8081 |
| Cr-Mn | 0.3538 | 1.3899 | 2.8507 |
| Cr-Al | 0.3794 | 1.2845 | 2.9952 |
| Fe-Ni | 0.4249 | 1.348 | 2.823 |
| Fe-Mn | 0.3507 | 1.3529 | 2.8655 |
| Fe-Al | 0.3761 | 1.2475 | 3.0124 |
| Ni-Mn | 0.355 | 1.3659 | 2.834 |
| Ni-Al | 0.3807 | 1.2605 | 3.028 |
| Mn-Al | 0.3142 | 1.2654 | 3.0235 |

It should be noted that, similar to the potential developed in [13], this potential is designed to describe multicomponent alloy, where most neighbours of a given atom are of different types. As a result, it does not always accurately capture the behaviour of pure elements.

The cutoff distance, $r_c$ was set to $2.5d$, where $d$ is the distance between the two nearest neighbouring atoms. For the FCC lattice of the CoCrFeNiMn alloy, with a lattice parameter of approximately 3.6 Å, this corresponds to a cutoff distance of 6.3 Å, encompassing up to the sixth nearest neighbour of each atom. Such a long-range cutoff is necessary because short-range pairwise potentials cannot effectively model the compositional randomness of the CoCrFeNiMn HEA, as they fail to capture the statistical variations in individual atomic neighbourhoods [13].

## 3. Results and discussion

During simulation 50000 atoms were deposited to the Al substrate, resulting in a film thickness of approximately 61 Å (Fig. 2). Each incident atomic type was present in a ratio of 20 percent. Based on the simulation results, it was established that during the initial stages of film growth, due to the high energy of 10 eV, incident atoms penetrate the interphase zone and diffuse into the Al substrate, creating defects (Fig.2a). Simultaneously, aluminium atoms diffuse into the growing film, altering its elemental composition. According to OVITO's polyhedral template matching analysis with an RMSD cutoff of 0.1, the film begins to crystallize after approximately 10 ns of simulation time, at a thickness of around 8 Å. During the initial stage of deposition, the film predominantly exhibits a BCC structure, likely due to its high aluminium content (Tab. 3, Fig. 3a).

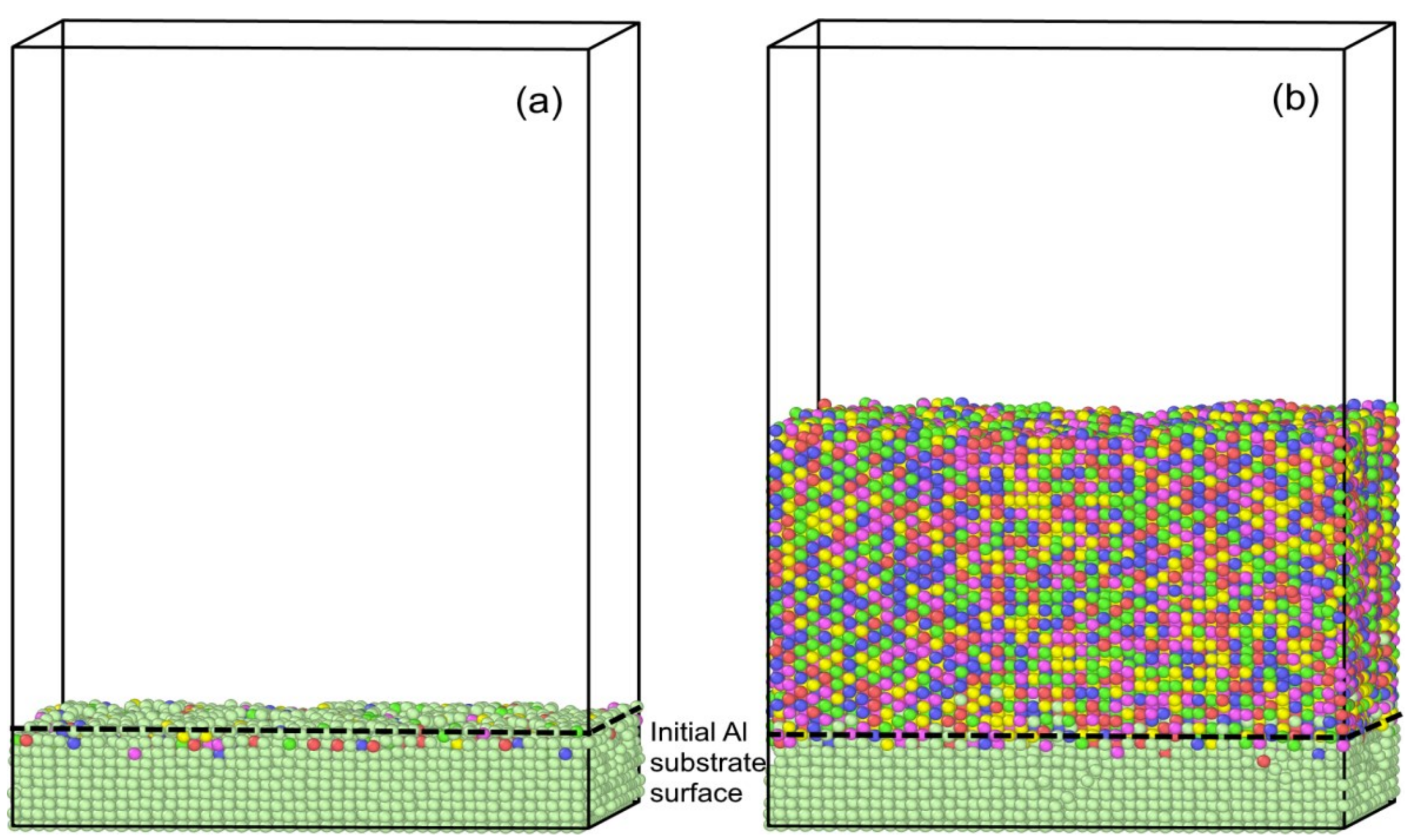

**Figure 2.** Deposited CoCrFeNiMn thin film shown at 1 ns (a) and 100 ns (b) of simulation time: ● – Al, ● – Co, ● – Cr, ● – Fe, ● – Ni, ● – Mn.

**Table 3.** Elemental composition of the simulated thin film (at. %).

| Simulation time | Al | Co | Cr | Fe | Ni | Mn |
|---|---|---|---|---|---|---|
| 10 ns | 11.3 | 16.6 | 17.6 | 17.9 | 16.9 | 19.7 |
| 20 ns | 4.8 | 18.5 | 19.0 | 19.1 | 18.7 | 19.9 |
| 30 ns | 3.0 | 19.1 | 19.4 | 19.4 | 19.2 | 19.9 |
| 40 ns | 2.2 | 19.3 | 19.5 | 19.6 | 19.4 | 20 |
| 50 ns | 1.7 | 19.5 | 19.6 | 19.7 | 19.5 | 20 |
| 60 ns | 1.5 | 19.6 | 19.6 | 19.7 | 19.6 | 20 |
| 70 ns | 1.2 | 19.6 | 19.7 | 19.8 | 19.7 | 20 |
| 80 ns | 1.1 | 19.7 | 19.7 | 19.8 | 19.7 | 20 |
| 90 ns | 0.9 | 19.7 | 19.8 | 19.8 | 19.8 | 20 |
| 100 ns | 0.9 | 19.7 | 19.8 | 19.8 | 19.8 | 20 |

As demonstrated in studies [31-34], the addition of Al to the CoCrFeNiMn HEA initiates the formation of a BCC phase. According to data from [32], the BCC phase begins to appear when the aluminium content exceeds 8 at.%, and when the aluminium content exceeds 16 at.%, the alloy consists entirely of the BCC phase. Then, the content of FCC and HCP phases begins to increase (Fig.3b), and after deposition is complete (100 ns of simulation), the simulated film contains an FCC phase, an HCP phase, a BCC phase, and an indefinite phase (Fig.3c, Fig4a). The latter, as indicated by the analysis of the radial distribution of atoms, exhibits an amorphous structure. It should be noted that Al atoms predominantly belong to the BCC phase only near the substrate surface. Atoms that have moved higher due to diffusion mainly belong to the FCC phase (Fig.3d). However, despite this, a small amount of the BCC phase is retained in the upper part of the film, mainly near the boundaries of the model. Table 3 presents the composition of the simulated film at various stages of deposition as well as its final composition. The configuration of individual grains in the polycrystalline microstructure of the deposited film, determined using OVITO grain segmentation analysis, is presented in Fig.4b.

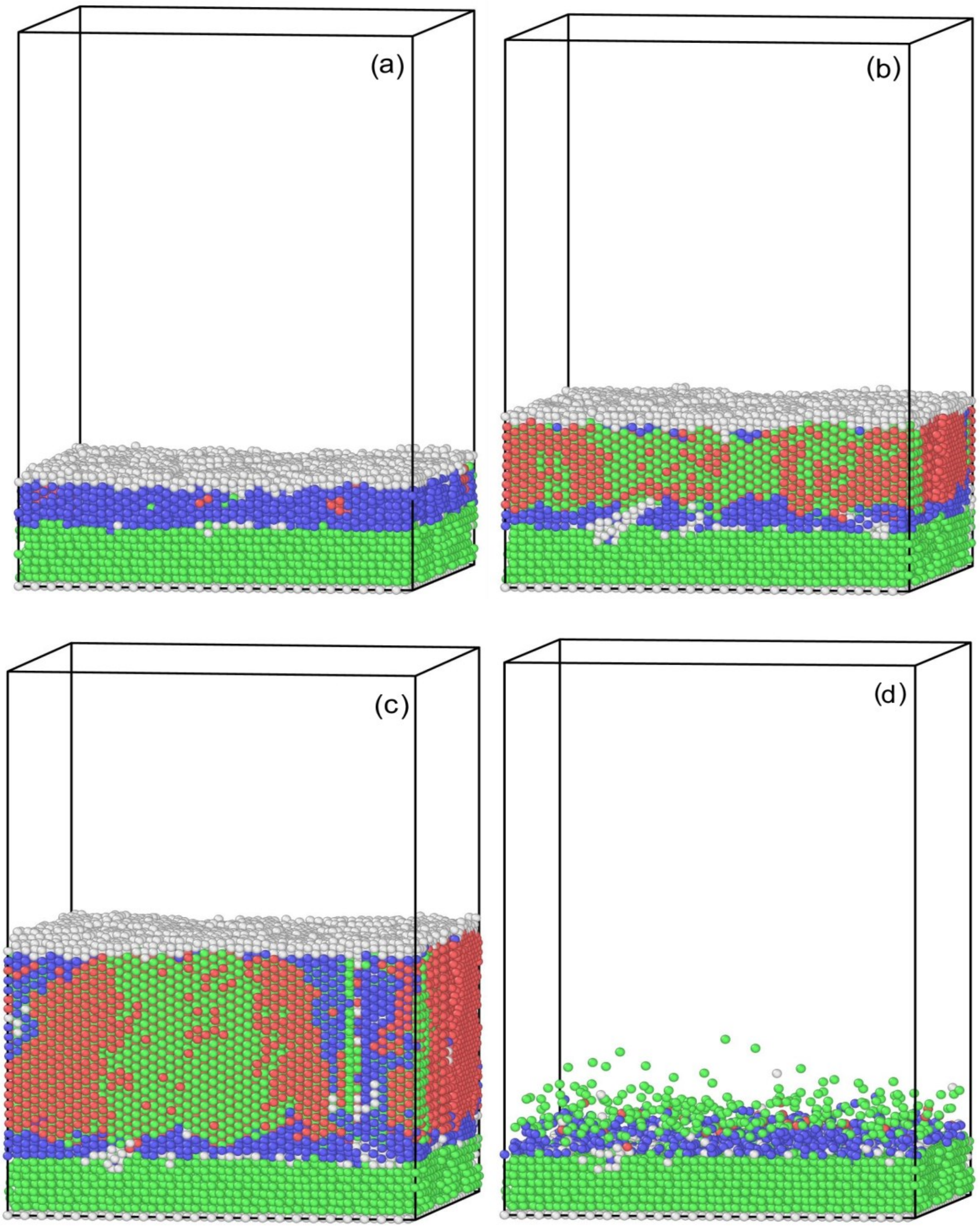

**Figure 3.** Structural evolution of CoCrFeNiMn thin film shown at 20 ns (a), 50 ns (b), and 100 ns (c) of simulation time; distribution of the Al atoms (d): ● – FCC, ● – BCC, ● – HCP, ● – amorphous phase.

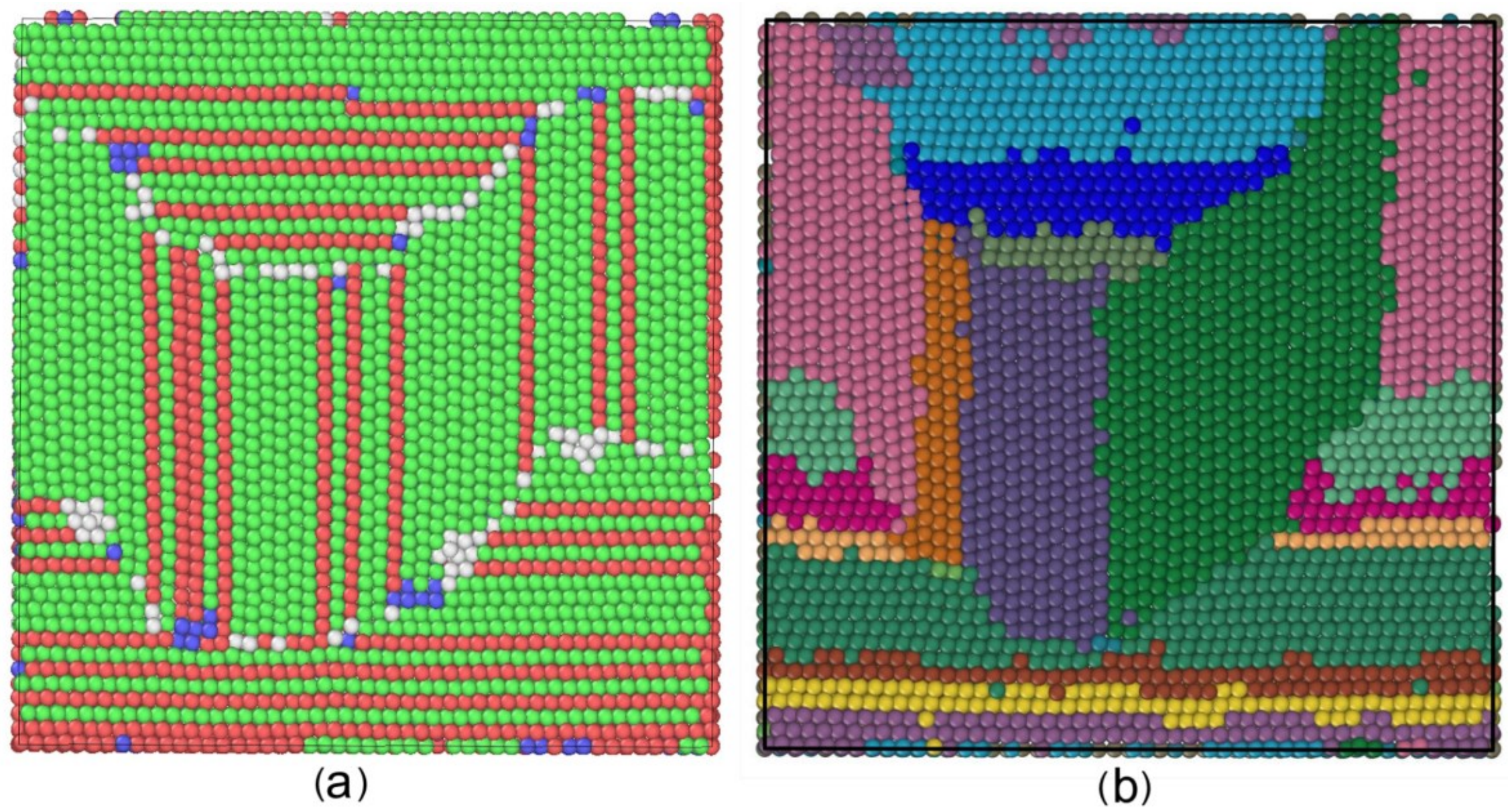

**Figure 4.** Structure of the CoCrFeNiMn thin film: (a) sliced along the substrate plane (top view), (b) result of OVITO grain segmentation analysis.

Determined from the analysis of RDF curves of simulated CoCrFeNiMn film (Fig.5) lattice parameters are: FCC $a$ = 3.610 Å, BCC $a$ = 2.906 Å, HCP $a$ = 2.524 Å, which is in good agreement with the lattice parameters of bulk samples and thin films of CoCrFeNiMn HEA obtained from experiments and DFT calculations [13,34]. It should be noted that the RDF curves were calculated excluding the atoms of the Al substrate but including the Al atoms that diffused into the film. Therefore, the slight increase in the calculated lattice parameters can be attributed to the addition of Al [33]. The peculiarities of the arrangement of atoms of the hexagonal close-packed (HCP) phase led us to the conclusion that this phase is a system of intrinsic and multilayer stacking faults in the lattice of the FCC phase. It also contains coherent twin boundaries.

It should also be noted that the HCP phase, according to the results of ab initio modelling, is one of the stable phases in the CoCrFeNiMn HEA [35]. In experiments, it was observed at high pressures [36-38], but local stresses that should appear during high-rate film deposition by atoms with high incident energy should facilitate the appearance of HCP phase in simulation. As for the amorphous phase, its appearance can be explained by the high cooling rate at which some of the atoms do not have enough time to rearrange and create a crystalline phase. It is worth noting that the content of the amorphous phase was significantly lower (17.9%) compared to simulations using the MEAM potential, where it reached 42.5% [19].

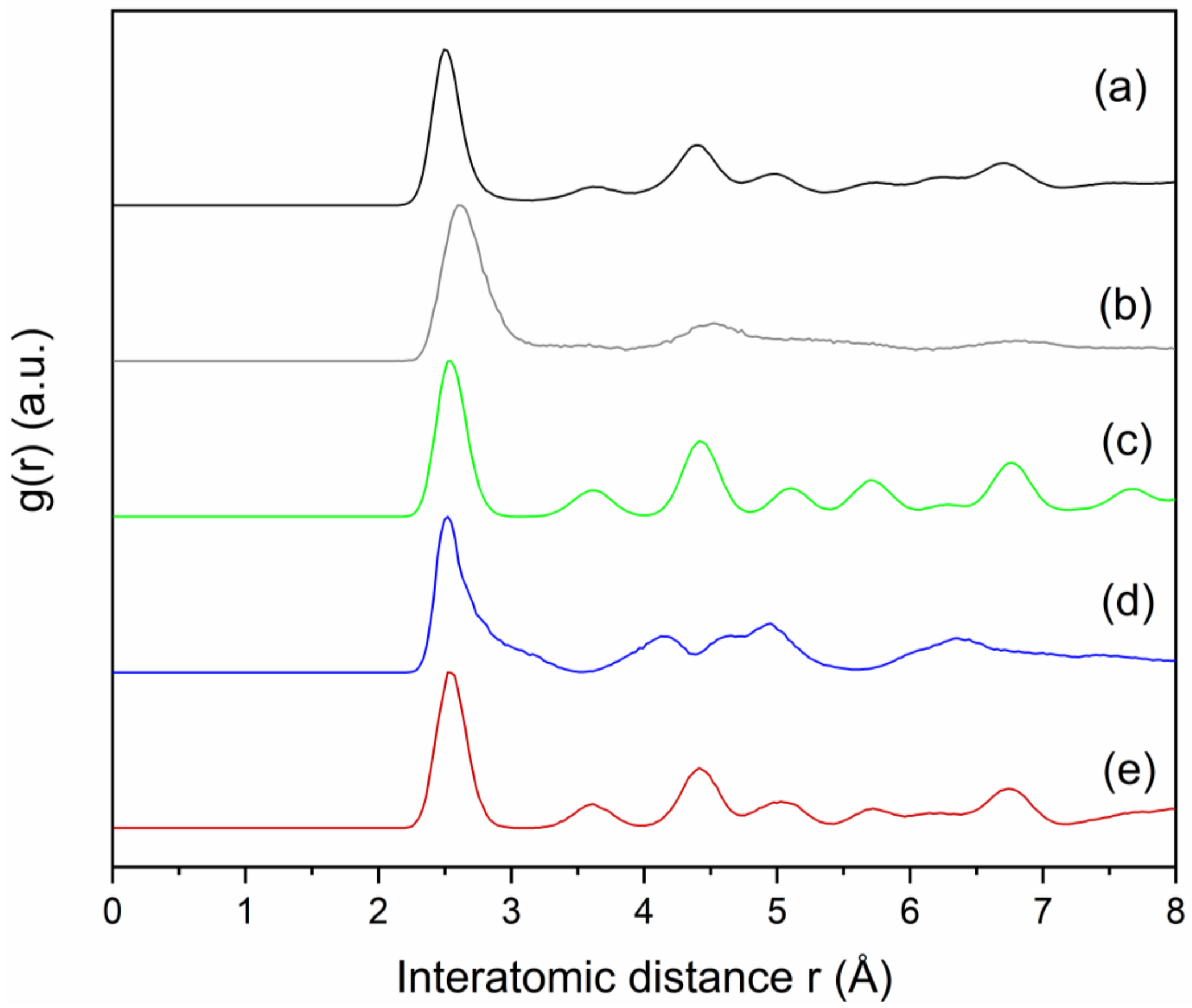

**Figure 5.** RDF curves of the CoCrFeNiMn thin film: (a) entire film, (b) amorphous phase, (c) FCC phase, (d) BCC phase, and (e) HCP phase.

## 4. Conclusions

Classical molecular dynamics simulations were performed to describe the growth and crystallization processes of thin CoCrFeNiMn HEA films. A set of Morse potentials was developed to model all interactions within the CoCrFeNiMn HEA, as well as between the HEA and the aluminium substrate. The structural characteristics and phase composition of the films were analysed using polyhedral template matching and radial distribution function (RDF) methods. According to the simulation results, the growth of CoCrFeNiMn films involves the formation of BCC, FCC, and HCP phases, along with an amorphous structure. The estimated lattice parameters for the BCC and FCC phases align well with experimental values for the CoCrFeNiMn HEA with the addition of aluminium. Thus, the developed interatomic potential is suitable for modelling the random multicomponent high-entropy CoCrFeNiMn alloy.

## Disclosure statement

No potential conflict of interest was reported by the author(s).

## ORCID

*O.I. Kushnerov* http://orcid.org/0000-0002-9683-2041

*S.I. Ryabtsev* http://orcid.org/0000-0002-2889-5278

*V.F. Bashev* http://orcid.org/0000-0002-3177-0935